\documentclass[conference,letterpaper]{IEEEtran}

\usepackage[utf8]{inputenc} 
\usepackage[T1]{fontenc}
\usepackage{url}
\usepackage{ifthen}
\usepackage{cite}
\usepackage[cmex10]{amsmath} 
\usepackage{algorithmic}
\usepackage{algorithm}
\usepackage{array}

 \usepackage[caption=false,font=normalsize,labelfont=sf,textfont=sf]{subfig}
\usepackage{tikz}
\usepackage{accents}
\usetikzlibrary{arrows.meta}
\usepackage{amssymb}
\usepackage{dsfont}
\usepackage[letterpaper, left=0.64in, right=0.64in, top=0.75in, bottom=0.5in, includefoot]{geometry}

\newcommand\munderbar[1]{%
  \underaccent{\bar}{#1}}

\newcommand{\xrseq}{\munderbar{X}}
\newcommand{\xrealiz}{\munderbar{x}}

\newcommand{\yrseq}{\munderbar{Y}}
\newcommand{\yrealiz}{\munderbar{y}}

\newcommand{\blocklen}{n}

\newcommand{\lrseq}{\munderbar{L}}
\newcommand{\lrealiz}{\munderbar{l}}
\newcommand{\lrate}{R_L}
\newcommand{\lbits}{\blocklen\lrate}

\newcommand{\krseq}{\munderbar{K}}
\newcommand{\krealiz}{\munderbar{k}}
\newcommand{\krate}{R_0}
\newcommand{\kbits}{\blocklen\krate}

\newcommand{\jrseq}{\munderbar{J}}
\newcommand{\jrealiz}{\munderbar{j}}
\newcommand{\jbits}{\blocklen\jrate}
\newcommand{\jrate}{R}

\newcommand{\xalph}{\munderbar{\mathcal{X}}}
\newcommand{\yalph}{\munderbar{\mathcal{Y}}}
\newcommand{\lalph}{\munderbar{\mathcal{L}}}
\newcommand{\kalph}{\munderbar{\mathcal{K}}}
\newcommand{\jalph}{\munderbar{\mathcal{J}}}

\newcommand{\numsamp}{N_{s}}

\newcommand{\targetJoint}{Q_{\xrseq \yrseq}}
\newcommand{\targetCond}{Q_{\yrseq | \xrseq}}
\newcommand{\targetInput}{Q_{\xrseq}}

\newcommand{\estTargetJoint}{\widehat{Q}_{\xrseq \yrseq}}

\newcommand{\predictionJoint}{P_{\xrseq \yrseq}}

\newcommand{\neuralClassifier}{P^{\theta}_{\yrseq|\xrseq\krseq\lrseq}}

\newcommand{\Klower}{k_{b-1}}
\newcommand{\Kupper}{k_{b}}
\newcommand{\Llower}{l_{b-1}}
\newcommand{\Lupper}{l_{b}}

\newcommand{\ybins}{B^{(\yrealiz)}}
\newcommand{\kbins}{B^{(\krealiz)}}
\newcommand{\lbins}{B^{(\lrealiz)}}

\newtheorem{theorem}{Theorem}

\newtheorem{definition}{Definition}

\begin{document}

\title{Deep Randomized Distributed Function Computation (DeepRDFC): Neural Distributed Channel Simulation}

\author{%
  \IEEEauthorblockN{Didrik~Bergström and Onur~Günlü}
  \IEEEauthorblockA{Information Theory and Security Laboratory (ITSL), Link{\"o}ping University, Sweden\\
                    Email: \{didrik.bergstrom, onur.gunlu\}@liu.se}
}%

\maketitle

\begin{abstract} 
The randomized distributed function computation (RDFC) framework, which unifies many cutting-edge distributed computation and learning applications, is considered. An autoencoder (AE) architecture is proposed to minimize the total variation distance between the probability distribution simulated by the AE outputs and an unknown target distribution, using only data samples. We illustrate significantly high RDFC performance with communication load gains from our AEs compared to data compression methods. Our designs establish deep learning-based RDFC methods and aim to facilitate the use of RDFC methods, especially when the amount of common randomness is limited and strong function computation guarantees are required.
\end{abstract}

\section{Introduction}
\label{sec:i-intro}
\IEEEPARstart{I}{n} conventional systems, data is communicated as arbitrary bit sequences without considering the conveyed meaning. This deficiency has motivated \textit{semantic communications} \cite{GunduzSemantic, PetarSemantic}, where the quality measure of transmission depends on the semantics \cite{GunduzBITS}. For example, transmitting objects and their relative positions, rather than raw pixel arrays, exemplifies semantic communication \cite{GunduzBITS}. As the semantics of data can be modeled as a function of the data, the semantic communication paradigm can be defined as a remote (hidden) source coding problem where the transmitted information is a function of the data, while the function output is not available to the transmitter \cite[pp.~78]{BergerBook}, \cite[pp.~118]{Csiszarbook}, \cite{DobrushinRemote}. This formulation extends beyond classical lossy coding by employing semantic-driven fidelity metrics \cite{OnurWIFS2024,GustafISIT2025,OnurEUCNC}.

We consider a generalization of remote source coding that employs a \emph{randomized} functional mapping for distributed function computation. In this setting, the receiver's computation relies on randomized transformations of the transmitter's data, thus defining a \emph{randomized distributed function computation (RDFC)} framework, introduced in \cite{OnurWIFS2024}. RDFC unifies a large set of important problems that leverage semantic communications. Example RDFC applications include machine learning-based data compression methods with generative models \cite{HavasiMRC,Flamich2020}, federated learning (FL) with side information \cite{Isik2024adaptive}, transform coding methods \cite{Agustsson2020,CheukQuant}, lossy compression methods with realism constraints (such as visual perception) \cite{Theisrealism,Blau2019}, and compression methods used as differential privacy mechanisms \cite{KairouzLDP,CheukDP}. These applications aim to synthesize data according to a target joint probability distribution, which broadly corresponds to controlling the correlation between data samples to satisfy a constraint requiring a controlled randomization step.
Minimizing the communication load required for RDFC is defined as a \emph{coordination problem} \cite{CuffChannelSynthesis,ReverseShannon,GerhardChannelSimulation,HarshaOriginalOneShot}, also known as distributed channel simulation/synthesis.

We focus on strong coordination \cite{CuffChannelSynthesis}, imposing joint typicality constraints on each function computation. Such a stringent constraint is vital to ensure a performance guarantee for each computation instance, unlike previous methods whose guarantee is for the average case over all computation instances, where the latter provides empirical coordination. Moreover, unlike computations of deterministic functions, common randomness shared between transmitter and receiver can significantly reduce the communication load for RDFC \cite{MACChannelSimulation}. Such uniformly distributed common randomness can be obtained, e.g., by using physical unclonable functions \cite{MyPUFLowComplexity}. For instance, if encoder-decoder pairs that apply RDFC methods are used to achieve local differential privacy, the communication load is reduced by up to 214 times compared to adding noise for the same purpose \cite{OnurWIFS2024}. Thus, in this work, we propose deep learning methods to design low-complexity encoder-decoder pairs for the RDFC framework. There are a few code constructions proposed for empirical coordination problems, such as in \cite{PolarCoordinationAaron}, designed for binary symmetric channels (BSCs), and existential designs for strong coordination, such as in \cite{MatthieuRemiPolarCoordination}. These existing designs do not provide constructive code designs for RDFC and do not leverage the high performance of deep neural networks, illustrated to surpass classical code constructions for reliable communications \cite{HoydisDNN}.

Our summary of the main contributions is as follows: 
\begin{itemize}
    \item We develop a general constructive design of autoencoders (AEs) for RDFC in a discrete setting. We propose a suitable loss function for RDFC and algorithms to generate training data that provide high RDFC performance. 
    \item We provide an AE architecture and technical insights to facilitate the training and application of AEs in RDFC. We motivate our choices of layer activation functions, training parameters, and the need for a vector quantizer layer that determines the RDFC rate for our setting.
    \item We demonstrate the feasibility of our design by simulating a BSC in a distributed computation setting. We also illustrate the impact of the amount of available common randomness on the RDFC performance.
\end{itemize}

\emph{Notation}: We denote random variables as capital letters \(X\) with realizations \(x\) from an alphabet \(\mathcal{X}\), and multivariate random variables of arbitrary dimensionality as underlined capital letters \(\xrseq\) with realizations \(\xrealiz\) from an alphabet \(\xalph\). Denote a binning set \(\ybins = \{\ybins_1, \ybins_2, \ldots\}\) as a collection of disjoint exhaustive sets of \(\yrealiz\), where set \(\ybins_b\) is indexed by \(b\). A similar notation is used for the binning set \(\kbins(\cdot) = \{\kbins_1(\cdot), \kbins_2(\cdot), \ldots\}\), where a function \(\kbins_b(\cdot)\) that maps an input to a defined set of realizations \(\krealiz\) is indexed by \(b\). Denote a Markov chain as \(X-U-Y\), i.e., $P_{XUY}=P_{XU}P_{Y|U}$. The total variation distance (TVD) is defined as $\left\|P - Q\right\|_{TV} \triangleq \frac{1}{2}\sum_{a\in \mathcal{A}} |P(a) - Q(a)|$, where \(P\) and \(Q\) are probability mass functions over a countable set $\mathcal{A}$. We denote the \(\ell^2\) norm as \(\|\!\cdot\!\|_{2}\). The loss function categorical cross-entropy (CCE) is defined as \(\text{CCE}(\yrealiz, \widehat{\yrealiz}) \triangleq - \yrealiz\ln{\widehat{\yrealiz}}\), where \(\yrealiz\) is a one-hot encoded ground truth observation, and \(\widehat{\yrealiz}\) is the AE output given inputs \((\xrealiz, \krealiz, \lrealiz)\). A BSC with crossover probability \(p\) is denoted as \(\text{BSC}(p)\). We denote a uniform distribution over the support of \(\krseq\), for example, as \(U_{\krseq}\). 

In Section~\ref{sec:systemmodel}, we introduce the RDFC framework and review AEs. In Section~\ref{sec:iii-autoencoder-dcs}, we propose AEs designed for the RDFC framework. In Section~\ref{sec:iv-results}, we present and discuss the results of the experiments performed using our AE designs. Finally, we conclude the paper in Section~\ref{sec:v-conclusion}.

\section{System Model and Review of AEs}\label{sec:systemmodel}
\subsection{Distributed Channel Simulation}
\label{sec:ii-dcs}

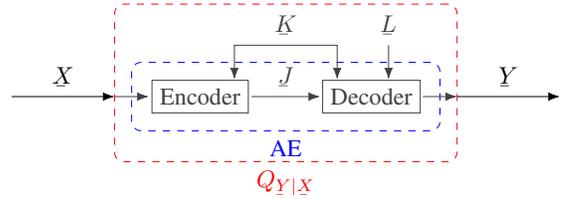
\begin{figure}[!t]
    \centering
    \resizebox{0.85\linewidth}{!}{
    \begin{tikzpicture}
    \draw[-Latex] (-4,0) -> (-2.5,0) node[pos=0.5, above]{\(\xrseq\)};
     \draw[-Latex, darkgray] (-2.5,0) -> (-2,0);
    \draw (0,0.2) node[minimum height=2.3cm,minimum width=5cm,draw, red, dashed, rounded corners, label={[name=label node, red]below:\(\targetCond\)}] {};
    \draw (0,0) node[minimum height=1cm,minimum width=4.5cm,draw, blue, dashed, rounded corners, label={[name=label node, blue]below:AE}] {};
    \draw (-1.25,0) node[minimum width=1cm, draw, darkgray] {Encoder};
    \draw[-Latex, darkgray] (-0.5,0) -> (0.5,0) node[pos=0.5, above, darkgray]{\(\jrseq\)};
    \draw[Latex-Latex, darkgray] (-0.75,0.25) -- (-0.75,0.75) -- (0.75,0.75) node[pos=0.5, above, darkgray]{\(\krseq\)} -- (0.75,0.25);
    \draw[-Latex, darkgray]  (1.5,0.75) node[above] {\(\lrseq\)} -- (1.5,0.25);
    \draw (1.25,0) node[minimum width=1cm, draw, darkgray] {Decoder};
     \draw[-Latex, darkgray] (2,0) -> (2.5,0);
    \draw[-Latex] (2.5,0) -> (4,0) node[pos=0.5, above]{\(\yrseq\)};
\end{tikzpicture}
}
    \caption{Distributed channel simulation system model, implemented using our AE designs.}
    \label{fig:dcs-ae-model}
\end{figure}

Consider the distributed channel simulation problem with target probability distribution \(\targetCond\) depicted in Fig.~\ref{fig:dcs-ae-model} that we aim to synthesize using an encoder-decoder pair. The encoder observes a discrete independent and identically distributed (i.i.d.) source input sequence \(\xrseq \sim \targetInput\in \mathcal{X}^n\) as well as uniformly distributed and independent common randomness \(\krseq \in \kalph=\![1\!:\!2^{\blocklen \krate}]\), the latter of which is available to both encoder and decoder. These inputs are encoded into an index \(\jrseq \in \jalph=[1\!:\!2^{\blocklen\jrate}]\) that is transmitted through a noiseless channel to the decoder. The decoder aims to output \(\yrseq\in \mathcal{Y}^{\blocklen}\) given the index \(\jrseq\), common randomness \(\krseq\), and uniformly distributed and independent local randomness \(\lrseq \in \lalph =[1\!:\!2^{\blocklen \lrate}] \) such that \((\xrealiz, \yrealiz)\sim \targetJoint=\targetInput\targetCond\) for all $(\xrealiz, \yrealiz)\in \mathcal{X}^n\times\mathcal{Y}^n$. We denote the encoder-decoder behavior as \(P_{\yrseq \jrseq|\xrseq \krseq \lrseq}\), which together with an input probability distribution results in the \emph{synthesized joint distribution}

\begin{equation}
    \predictionJoint(\xrealiz, \yrealiz) = \!\!\!\!\!\! \sum_{\substack{\jrealiz, \krealiz,\lrealiz\in \jalph\times \kalph\times \lalph}}\!\!\!\!\!  U_{\krseq} \cdot U_{\lrseq} \cdot P_{\xrseq}(\xrealiz) \cdot P_{\yrseq \jrseq|\xrseq \krseq \lrseq}(\yrealiz, \jrealiz |\xrealiz, \krealiz,\lrealiz). 
\end{equation}

Next, we define the RDFC rate region for the distributed channel synthesis problem. 

\begin{definition}
    A rate triple \((\jrate, \krate, \lrate)\) is achievable for \(\targetJoint\) if there exists \(\blocklen\!\geq\!1\) and an encoder-decoder pair with the synthesized joint distribution \(\predictionJoint\) such that
    \begin{align}
       \lim_{\blocklen \to \infty} \big\| \predictionJoint - \prod_{i=1}^n Q_{XY} \big\|_{TV} = 0.\label{eq:TVDasympt}
    \end{align}
The RDFC rate region \(\mathcal{R}\) is the closure of the set of all achievable rate triples. \hfill $\lozenge$
\end{definition}

Below, we provide the RDFC rate region \(\mathcal{R}\) for the distributed local synthesis problem with local randomness.

\begin{theorem}[{\hspace{1sp}\cite[Eq. (42)]{CuffChannelSynthesis}}]
The RDFC rate region $\mathcal{R}$ is 
\begin{equation} \label{th:rate-region}
    \mathcal{R} \triangleq 
    \bigcup_{\substack{P_{UXY} : P_{XY} = Q_{XY},\\X-U-Y}} \left\{
    \begin{aligned} 
        &(\jrate, \krate, \lrate) : \jrate  \geq I(X;U), \\
        &\krate + \jrate \geq I(X,Y;U),\\
        &\lrate \geq H(Y|U)
    \end{aligned}
    \right\}.
\end{equation}
It suffices to have \(|\mathcal{U}| \leq |\mathcal{X}||\mathcal{Y}|+2\).
\end{theorem}

In this point-to-point scenario, two extremes characterize the RDFC rate region, given that there is enough local randomness. Without common randomness, i.e., $R_0=0$, the minimal communication rate is equal to Wyner's common information (WCI) $C(X;Y)=\inf_{U:X-U-Y} I(X,Y;U)$\cite{WCI}. With large enough common randomness $\krate$, the distribution rate $\jrate$ can be decreased to $I(X;Y)$.

\subsection{Review of AEs}
\label{sec:ii-ae-construction}

An AE is a feed-forward neural network consisting of an encoder that outputs an index \(\jrealiz = \text{Enc}(\xrealiz)\) given an input \(\xrealiz\), and a decoder that receives \(\jrealiz\) and reconstructs the message \(\widehat{\yrealiz} = \text{Dec}(\jrealiz)\) \cite{AutoencoderDimReduction}. The AE is trained by running the above step (forward propagation/inference) and then computing the loss according to a loss function \(\mathcal{L}(\yrealiz, \widehat{\yrealiz})\). In the following step (backpropagation), the chain rule is applied to compute the gradients of the loss function with respect to all network parameters, starting from the output layer and moving backward through the network. These gradients are then passed to an optimizer that updates the network parameters, thereby reducing the loss on the next iteration.

In our setting, the AE takes the source input \(\xrseq\) and common- and local randomness \(\krseq\) and \(\lrseq\) as inputs, and outputs \(\yrseq\). The index \(\jrseq\) is contained within the AE and automatically adapts itself, given some training set \(\{\xrealiz_i, \yrealiz_i, \krealiz_i, \lrealiz_i\}_{i=1}^{\numsamp}\), to improve the RDFC performance during training. We measure performance using the non-asymptotic TVD given in (\ref{eq:TVDasympt}).

\section{AE-based RDFC}
\label{sec:iii-autoencoder-dcs}
In this section, we first provide our proposed algorithms to generate training data and motivate our choice of the loss function for RDFC. Moreover, we describe our training and testing methodology, including the vector quantization layer. We use an AE to jointly optimize an encoder-decoder pair, as used in \cite{HoydisDNN}, for reliable communications, given only samples from the target distribution \(\targetJoint\).

Denote an AE as \(\neuralClassifier\), where \(\theta\) denotes the network parameters. Similarly, denote the synthesized joint distribution as
\begin{align}
    \predictionJoint(\xrealiz, \yrealiz) = \sum_{\krealiz,\lrealiz} P_{\xrseq} (\xrealiz) \cdot \neuralClassifier( \yrealiz |\xrealiz, \krealiz, \lrealiz) \cdot U_{\krseq} \cdot U_{\lrseq}.
\end{align}
The RDFC performance depends on the distribution of samples in the training set. Since the inputs \(\xrseq\), \(\krseq\), and \(\lrseq\) essentially determine the output \(\yrseq\), we must construct a training set that offers examples for learning. We provide this construction method in Algorithms~\ref{alg:KLBinning} and \ref{alg:TrainingDataGen}.

\begin{algorithm}[ht]
    \caption{Binning of K and L}\label{alg:KLBinning}
    \renewcommand{\algorithmicrequire}{\textbf{input:}}
    \renewcommand{\algorithmicensure}{\textbf{output:}}
    \renewcommand{\algorithmicreturn}{\textbf{return:}}
    \begin{algorithmic}[1]
        \REQUIRE Estimated joint target probability $\estTargetJoint$, \(|\kalph|\), \(|\lalph|\), bin-width \(\beta = 2^i,\, i=1,2,..., n\)
        
        \ENSURE Output bins \(\ybins\), and K/L-bins \(\kbins(\xrealiz)\) \\and \(\lbins(\xrealiz,\yrealiz)\)

        \STATE Let \(\ybins = \{\ybins_1, \ybins_2, ..., \ybins_{\frac{|\yalph|}{\beta}}\}\) be a partition of \(\yalph\) such that \(\ybins_b = \{(b-1)\beta,..,b\beta-1\}, b = 1,2,...,\frac{|\yalph|}{\beta}\) 
        \STATE Number of bins \(|\ybins| \gets \frac{|\yalph|}{\beta}\)

        \FORALL{$\xrealiz \in \xalph$}

            \STATE Initialize \(k_0 \gets -1\) 
            \FOR{$b = [1:|\ybins|]$}

                \STATE \(\Pr[\ybins_b | \xrseq = \xrealiz ] \gets \sum_{\yrealiz \in \ybins_b} \widehat{Q}_{\yrseq = \yrealiz|\xrseq=\xrealiz}\) \label{eq:k-constr-prob}
                
                \STATE Find \(\Kupper : \frac{\displaystyle \Kupper - \Klower}{\displaystyle |\kalph|} \approx  \Pr[\ybins_b | \xrseq = \xrealiz ]\)
                \STATE  \(\kbins_b(\xrealiz) \gets [\Klower+1:\Kupper]\)

            \ENDFOR
            \IF{not [allow empty K-bins]}
            \STATE \texttt{assert:} $\nexists \, b \in [1:|\ybins|]: \kbins_b(\xrealiz) = \varnothing$
            \ENDIF
            \STATE \texttt{assert:} \(\bigcup_{b} \kbins_b(\xrealiz) = \kalph\)
            \FOR{$b = [1:|\ybins|]$}

                \STATE \(\Pr[\yrseq = \yrealiz | \xrseq = \xrealiz, \ybins_b] \gets \frac{\displaystyle \widehat{Q}_{\yrseq = \yrealiz|\xrseq=\xrealiz}}{\displaystyle \Pr[\ybins_b | \xrseq = \xrealiz ]}\) \label{eq:l-constr-prob}
                \STATE Initialize \(l_0 \gets -1\)
                \FORALL{\(\yrealiz \in \ybins_b\)}
                    \STATE Find \(\Lupper : \frac{\displaystyle \Lupper - \Llower}{\displaystyle |\lalph|} \approx \Pr[\yrseq = \yrealiz | \xrseq = \xrealiz, \ybins_b]\)
                    \STATE \(\lbins(\xrealiz, \yrealiz) \gets [\Llower+1:\Lupper]\)
                \ENDFOR
                \STATE \texttt{assert:} \(\bigcup_{\yrealiz \in \ybins_b} \lbins(\xrealiz, \yrealiz)  = \lalph\)
            \ENDFOR
        \ENDFOR

    \end{algorithmic}
\end{algorithm}

\begin{table}
\begin{center}
\caption{General layout of AE, where for cases without common randomness, we set \(\kbits=0\) and remove layer 2.}
\label{tab:ae-layout}
\begin{tabular}{|c| c | c |c |}
\hline
\# & Layer & In-Layer&  Output Dim. \\
\hline
1 & Input \(\xrealiz\) & ---& \(2^{\blocklen}\) \\
\hline
2& Input \(\krealiz\)&  ---& \(\kbits\) \\
\hline
3&Concatenate &  1,2& \(4(2^{\blocklen} + \kbits)\) \\
\hline
4-6&Dense + ReLU &  3& \(4(2^{\blocklen} + \kbits)\) \\
\hline
7&Dense + Sigmoid & 6 & \(n-1\) \\
\hline
8&Vector Quantizer &  7 & \(n-1\)\\
\hline
9&Input \(\lrealiz\)&  --- & \(\lbits\) \\
\hline
10&Concatenate &  2,8,9& \(6(2^{\blocklen} + \kbits + \lbits)\) \\
\hline
11-15&Dense + ReLU &  10 & \(6(2^{\blocklen} + \kbits+ \lbits)\) \\
\hline
16&Dense + Softmax &  15 & \(2^{\blocklen}\) \\
\hline

\end{tabular}
\end{center}
\end{table}

\begin{algorithm}
    \caption{Training Data Generation}\label{alg:TrainingDataGen}
    \renewcommand{\algorithmicrequire}{\textbf{input:}}
    \renewcommand{\algorithmicensure}{\textbf{output:}}
    \renewcommand{\algorithmicreturn}{\textbf{return:}}
    \begin{algorithmic}[1]
        \REQUIRE Samples of channel input/outputs $\{\xrealiz_i, \yrealiz_i\}_{i=1}^{\numsamp}$ from $\targetJoint$, output bins \(\ybins\), and K/L-bins \(\kbins(\xrealiz)\), \(\lbins(\xrealiz, \yrealiz)\)
        
        \ENSURE Samples $\{\xrealiz_i, \yrealiz_i, \krealiz_i, \lrealiz_i\}_{i=1}^{\numsamp}$

        \FOR{$i = [1..\numsamp]$}
            \STATE \(\krealiz_i \gets \text{Uniformly sampled from } \kbins_b(\xrealiz_i),\, b:\yrealiz_i \in \ybins_b\)
            \STATE $\lrealiz_i \gets \text{Uniformly sampled from } \lbins(\xrealiz_i, \yrealiz_i)$
        \ENDFOR

    \end{algorithmic}
\end{algorithm}

Algorithm~\ref{alg:KLBinning} constructs output bins \(\ybins\), common-randomness bins \(\kbins(\xrealiz)\), and local randomness bins \(\lbins(\xrealiz, \yrealiz)\). The output bins \(\ybins\) partition \(\yalph\) into intervals of width \(\beta\), where intervals are indexed with \(b\). For a given input-output pair \((\xrealiz, \yrealiz)\), we identify the interval \(b\) that \(\yrealiz\) is in via \(\ybins\), and estimate the probability that the AE outputs some \(\yrealiz \in \ybins_b\) via \(\widehat{Q}_{\yrseq|\xrseq=\xrealiz}\) (line \ref{eq:k-constr-prob}). We construct \(\kbins_b(\xrealiz)\) so that \(|\kbins_b(\xrealiz)|/|\kalph|\) corresponds to this probability. We then estimate the probability that the AE outputs \(\yrealiz\) given the same fixed \(\xrealiz\) and the interval \(b\) (line \ref{eq:l-constr-prob}). We construct \(\lbins(\xrealiz, \yrealiz)\) so that \(|\lbins(\xrealiz, \yrealiz)|/|\lalph|\) approximates this probability. Multiplying lines \ref{eq:k-constr-prob} and \ref{eq:l-constr-prob} gives \(\widehat{Q}_{\yrseq=\yrealiz|\xrseq=\xrealiz}\), repeated for all \((\xrealiz, \yrealiz) \!\in\! \xalph \times \yalph\). 

Algorithm~\ref{alg:TrainingDataGen} samples realizations \((\krealiz_i, \lrealiz_i)\) uniformly randomly from the sets that are mapped to the samples \((\xrealiz_i, \yrealiz_i)\) via the aforementioned bins and concatenates the samples into a training set. Furthermore, we use one-hot encoded input-output vectors and the softmax activation function in the output layer to interpret the output as the AE's maximum likelihood prediction \(\widehat{\yrealiz}\) given \((\xrealiz, \krealiz, \lrealiz)\). We round the output with $\arg\!\max$ during inference to emulate the AE's intended hard decisions. The common randomness \(\krseq\) and local randomness \(\lrseq\) are encoded as binary sequences. We implement the AE in Keras/Tensorflow with the design given in Table~\ref{tab:ae-layout}.

\subsection{Loss Function for TVD Minimization}
\label{sec:iii-loss}
Using TVD as the loss function is intractable in this setting because the gradient descent tends to get stuck in local minima. Therefore, we use CCE as our loss function during the training phase. CCE is a suitable surrogate loss function to TVD due to (i) its differentiability and (ii) its equivalence to Kullback-Liebler divergence if the AE's inputs and outputs are one-hot encoded, which is the case in our design. Moreover, Kullback-Liebler divergence bounds TVD from above via Pinsker's inequality \cite{Elgamalbook}.

\subsection{Training and Testing}
\label{sec:iii-training}
As in \cite{HoydisDNN}, we use a large batch size (\(2^{14}\)) and a low learning rate. We use the ADAM optimizer with parameters \((lr\!=\!10^{-4},\, \beta_1\!=\!0.9,\, \beta_2\!=\!0.999)\). To adaptively lower the learning rate when the loss plateaus, we use a \textit{ReduceLROnPlateau} callback\cite{KerasReduceCallback} with parameters \((\)monitor=``loss", patience=\(1\), min\_delta=\(0.01)\). We use a set of samples \(\{\xrealiz_i, \yrealiz_i\}_{i=1}^{\numsamp}\) from the target joint distribution \(\targetJoint\) to compute the estimated distribution \(\estTargetJoint\) as a relative frequency distribution, where the total number of samples is \(\numsamp \!=\! 2^{26}\). We then construct the training set \(\{\xrealiz_i, \yrealiz_i, \krealiz_i, \lrealiz_i\}_{i=1}^{\numsamp}\) using Algorithms~\ref{alg:KLBinning} and \ref{alg:TrainingDataGen}. The AE \(\neuralClassifier\) is trained for \(20\) epochs to obtain 
$\theta^* : \underset{\theta}{\mathrm{argmin}}\, \sum_{i=1}^{\numsamp} -\yrealiz_i\log \widehat{\yrealiz_i}$, where \(\theta^*\) denotes the optimal values for the AE parameters \(\theta\), and \(\widehat{\yrealiz}_i\) is the AE's output given \((\xrealiz_i, \krealiz_i, \lrealiz_i)\) from the test set. 

Before testing, we create a separate \emph{test set} \(\{\xrealiz_i, \yrealiz_i, \krealiz_i, \lrealiz_i\}_{i=1}^{\numsamp}\) where input-output realizations \((\xrealiz_i, \yrealiz_i)\) are sampled from \(\targetJoint\) and common- and local randomness realizations are uniformly randomly sampled from their respective alphabets. We also compute an estimated joint target distribution \(\estTargetJoint\) for the test set, as above.

We test the AE \(\neuralClassifier\) by letting it make predictions \(\{\widehat{\yrealiz}_i\}_{i=1}^{\numsamp}\) given inputs \(\{\xrealiz_i, \krealiz_i, \lrealiz_i\}_{i=1}^{\numsamp}\) from the test set. We then apply the $\arg\!\max$ function to the predictions and compute the synthesized joint distribution \(\predictionJoint\) as a relative frequency distribution of observations made from the prediction set \(\{\xrealiz_i, \widehat{\yrealiz}_i\}_{i=1}^{\numsamp}\).
Finally, we compute the target joint distribution \(\targetJoint\), the testing TVD: \(\|\predictionJoint - \estTargetJoint\|_{TV}\) and the ground-truth TVD: \(\|\predictionJoint - \targetJoint\|_{TV}\).

\subsection{Vector Quantizer Layer}
\label{sec:iii-vq}

We constrain the RDFC rate \(\jrate\) by using a vector quantizer (VQ) layer, as in \cite{van2017neuralVQ}. The VQ-layer is initialized with a fixed embedding space \(\jalph = [\jrealiz_1, \jrealiz_2,...,\jrealiz_{2^{\blocklen\jrate}}]\), and we ensure that the VQ-layer has a preceding layer with an \(\blocklen\jrate\)-dimensional output \(\tilde{\jrealiz}\). The VQ-layer takes \(\tilde{\jrealiz}\) as input and outputs the closest discrete index as
\begin{align}\label{eq:vq-argmin-quantization}
    \jrealiz_i \in \jalph : i=\underset{i}{\mathrm{argmin}}\,\|\tilde{\jrealiz} - \jrealiz_i\|_2.
\end{align}
Since (\ref{eq:vq-argmin-quantization}) is not differentiable, the VQ-layer uses a straight-through estimator\cite{van2017neuralVQ} to pass the gradient from the decoder to the layer preceding the VQ-layer during backpropagation. We adapt the implementation in\cite{KerasVectorQuantizer} to our specific design.

\section{Experimental Results}
\label{sec:iv-results}
We consider a \(\text{BSC}(p)\) as our target conditional probability distribution \(\targetCond\) with a binary uniform i.i.d. source input sequence \(\xrseq \sim \targetInput\) and a binary output sequence \(\yrseq\) of blocklength \(\blocklen\). Both common randomness \(\krseq\) and local randomness \(\lrseq\) are uniformly random i.i.d. binary sequences. We consider an index \(\jrseq\) that is binary and of length \(\jbits\!=\! \blocklen -1\), which the VQ-layer imposes. We list the blocklengths, the other design parameters, and the results from our experiments in Table~\ref{tab:results}. The upper half of Table~\ref{tab:results} lists the experiments where only local randomness is available to the decoder, denoted as the \textit{LR-cases}. The lower half of Table~\ref{tab:results} (denoted as the \textit{LR+CR-cases}) lists the experiments where common randomness is available to both the encoder and decoder and local randomness is available to the decoder. We denote the test TVD \(\| \predictionJoint - \estTargetJoint\|_{TV}\) as \(\text{TVD}_T\) and the ground truth TVD \(\| \predictionJoint - \targetJoint\|_{TV}\) as \(\text{TVD}_G\), respectively.

\begin{table}
    \begin{center}
    \caption{TVD hyper-parameter data in LR-case and LR+CR-case.}
    \label{tab:results}
    \begin{tabular}{|c|c|c|c|c|c|c|}
    \hline
        \(\blocklen\) &  $\kbits$ & $\lbits$ & $\jbits$ & $p$ & \(\text{TVD}_T\) & \(\text{TVD}_G\)\\
        \hline
         $8$ & -- & $12$ & $7$ & $0.11$ &  \(0.575393\) & \(0.575175\) \\
        \hline
        $8$ & -- & $12$ & $7$ & $0.25$ & \(0.349378\) & \(0.348698\) \\
        \hline
        $10$ & -- & $15$ & $9$ & $0.11$ & \(0.553410\) & \(0.551960\)\\
        \hline
        $10$ & -- & $15$ & $9$ & $0.25$ & \(0.370869\) & \(0.366169\) \\
        \hline
        $8$ & -- & $16$ & $7$ & $0.11$ &  \(0.558904\) & \( 0.558633\) \\
        \hline
        $8$ & -- & $16$ & $7$ & $0.25$ & $0.346572$ & \(0.346019\) \\
        \hline
        $10$ & -- & $20$ & $9$ & $0.11$ & $0.524589$ & $0.523337$\\
        \hline
        $10$ & -- & $20$ & $9$ & $0.25$ & $0.366822$ & $0.363320$ \\ 
        \hline
        \hline
        $8$ & $16$ & $12$ & $7$ & $0.11$ & \(0.334202\) & \(0.333668\) \\  
        \hline
        $8$ & $16$ & $12$ & $7$ & $0.25$ & $0.074235$ & $0.072706$\\ 
        \hline     
        $10$ & $20$ & $15$ & $9$ & $0.11$ & \(0.442723\) & \(0.440944\) \\       
        \hline
        $10$ & $20$ & $15$ & $9$ & $0.25$ & \(0.135699\) & \(0.125946\)\\    
        \hline
        $8$ & $16$ & $16$ & $7$ & $0.11$ & \(0.334019\) & \(0.333433\) \\
        \hline
        $8$ & $16$ & $16$ & $7$ & $0.25$ &\(0.043022\) & \(0.040582\)\\
        \hline     
        $10$ & $20$ & $20$ & $9$ & $0.11$ & \(0.436145\) & \(0.434455\)\\   
        \hline
        $10$ & $20$ & $20$ & $9$ & $0.25$ & \(0.139957\) & \(0.130640\)\\
        \hline
    \end{tabular}
    \end{center}
\end{table}
As shown in Table~\ref{tab:results}, we select the common- and local randomness rates \(\krate\) and \(\lrate\) to be relatively large to ensure their achievability, asymptotically defined by (\ref{th:rate-region}). We set the RDFC rate \(\jrate\) to either \(7/8\) or \(9/10\), which suffices to demonstrate feasibility. Moreover, we observe from Table~\ref{tab:results} that the difference between the \(\text{TVD}_T\) and the \(\text{TVD}_G\) is small in both the LR- and LR+CR-cases, and \(\text{TVD}_G < \text{TVD}_T\) for all experiments. We expect \(\text{TVD}_G\) to be less than \(\text{TVD}_T\) in general, and the small differences between the two in Table~\ref{tab:results} indicate that our AE designs generalize well.

\begin{figure}[ht]
\captionsetup{farskip=0pt,nearskip=4pt}
\centering
    \subfloat[]{
\includegraphics[width=0.347\linewidth]{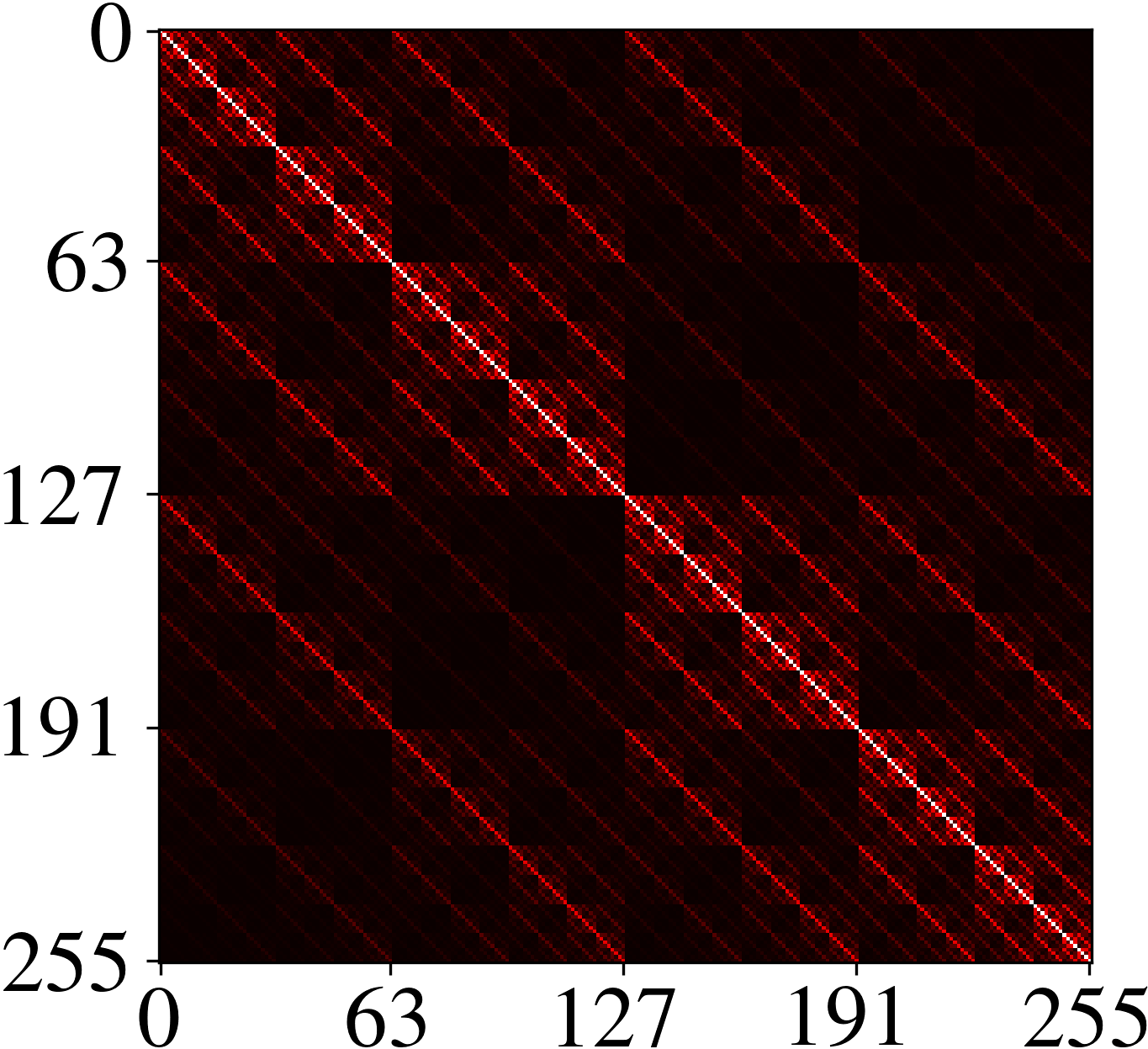}}%
\hfill
    \subfloat[]{
\includegraphics[width=0.304\linewidth]{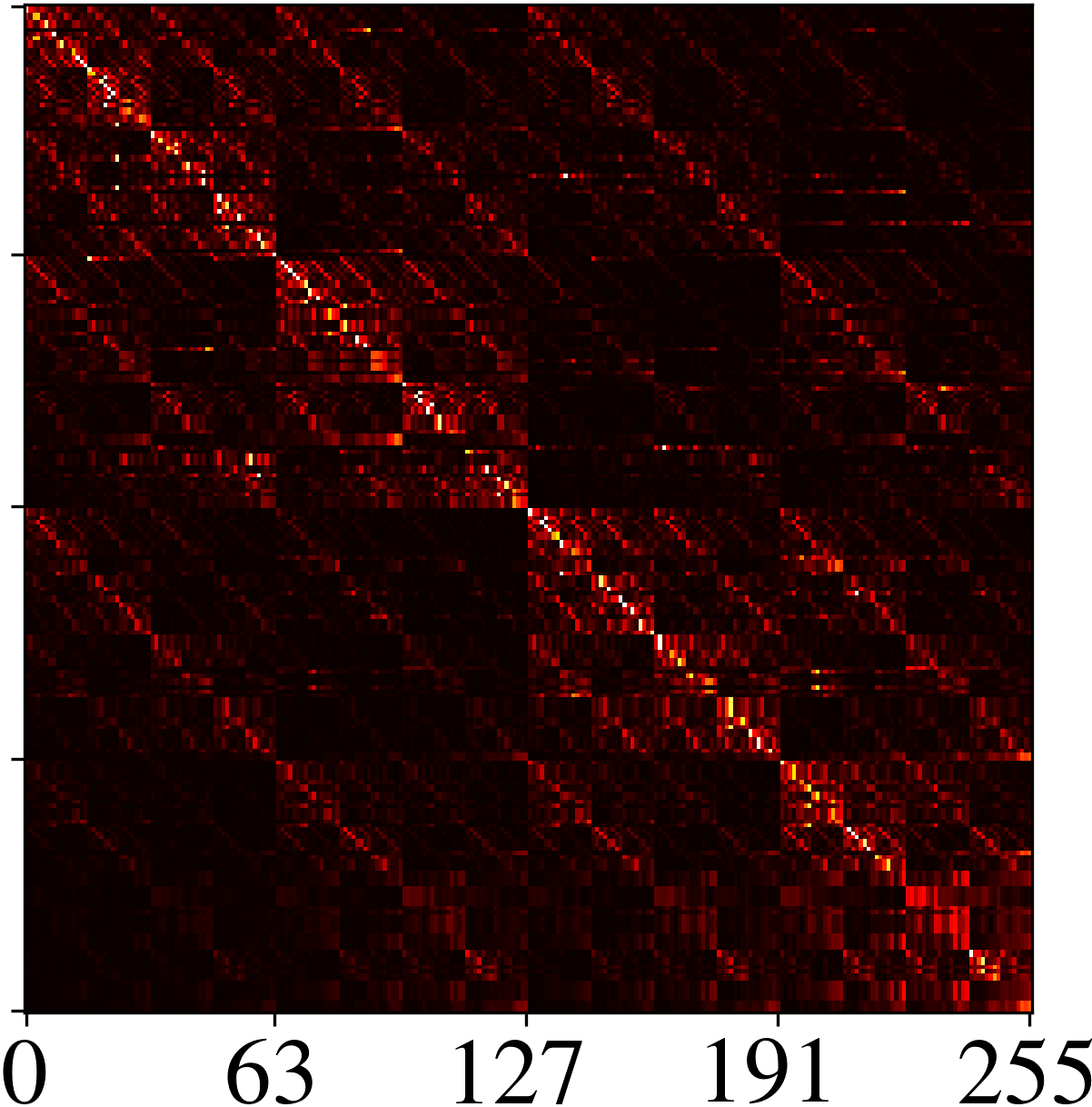}}%
\hfill
    \subfloat[]{
\includegraphics[width=0.304\linewidth]{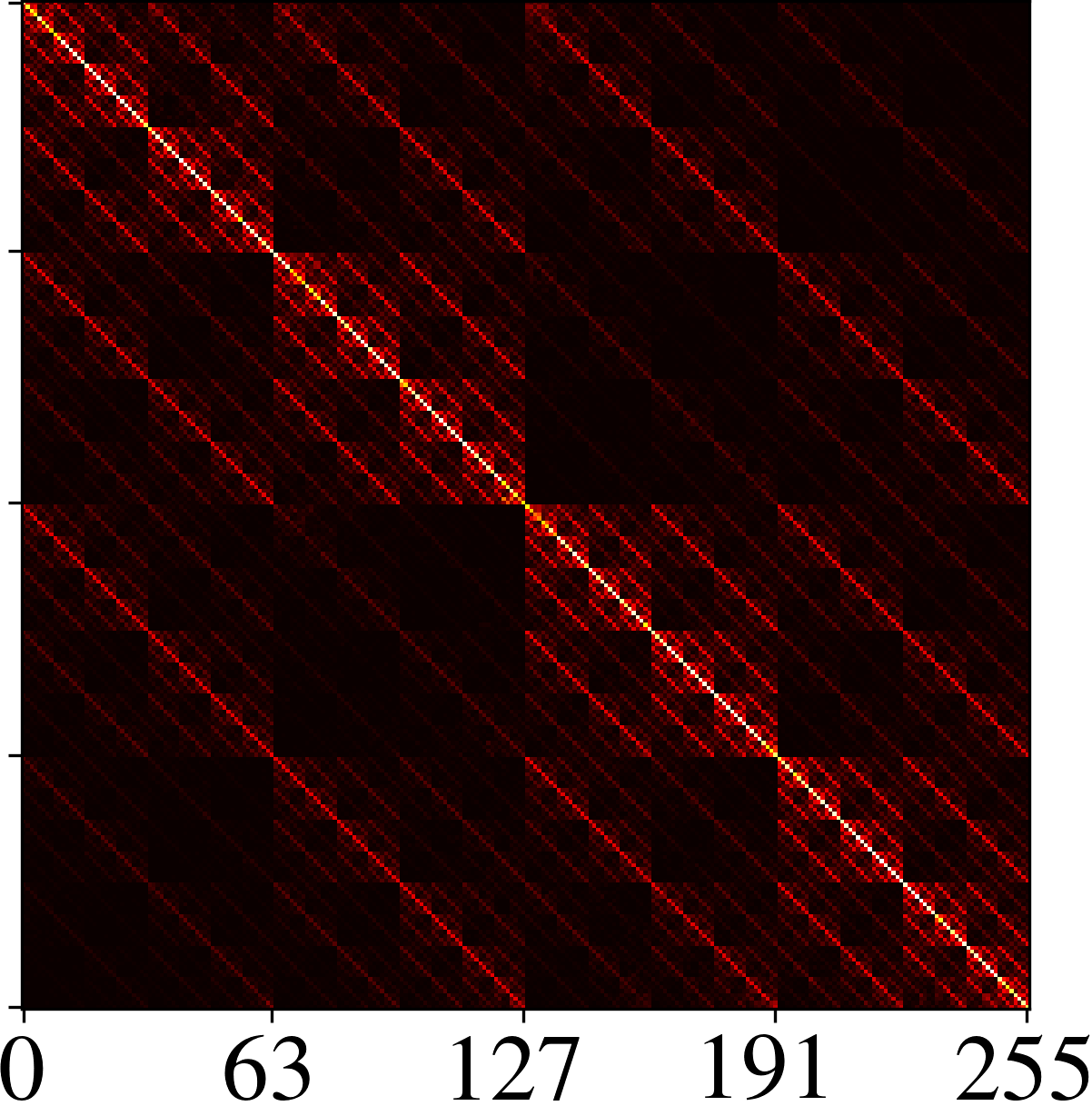}}%

\caption{Heatmaps for \(Q_{\protect\underaccent{\bar}{X}\protect\underaccent{\bar}{Y}}\) and \(P_{\protect\underaccent{\bar}{X}\protect\underaccent{\bar}{Y}}\)  of a \(\text{BSC}(p\!=\!0.25)\) and \(\blocklen\!=\!8\), where (a) is \(Q_{\protect\underaccent{\bar}{X}\protect\underaccent{\bar}{Y}}\); (b) is \(P_{\protect\underaccent{\bar}{X}\protect\underaccent{\bar}{Y}}\) with \(\lbits\!=\!16\), \(\kbits\!=\!0\), and \(\text{TVD}_T\approx0.35\); and (c) is \(P_{\protect\underaccent{\bar}{X}\protect\underaccent{\bar}{Y}}\) with \(\lbits\!=\!16\),\(\kbits\!=\!16\), and \(\text{TVD}_T\approx0.04\). The x- and y-axis correspond to \(\protect\underaccent{\bar}{x}\) and \(\protect\underaccent{\bar}{y}\), respectively.}
\label{fig:tvd-comparisons}
\end{figure}

We observe that the LR-cases achieve high RDFC performance, strictly improved by the LR+CR-cases. This improvement is expected as the AE learns to leverage the common randomness for distributed channel simulation for a fixed-rate index $\jrseq$. Furthermore, increasing \(\lrate\) also improves the RDFC performance, with the improvement being higher for larger crossover probabilities $p$. This result is also expected since the lower bound on \(\lrate\) in (\ref{th:rate-region}) increases with increasing $p$. We illustrate the similarities between joint distributions for different cases in Fig.~\ref{fig:tvd-comparisons}, where, e.g., the middle figure has a \(\text{TVD}_T\approx0.34\) yet maintains the target structure.

\begin{table}
    \begin{center}
         \caption{Sample sizes in relation to total number of combinations.}
    \label{tab:relative-samples}
  
    \begin{tabular}{|c|c|c|c|c|}
    \hline
        \(\blocklen\) &  $\kbits$ & $\lbits$  &\(T\) &  \(\numsamp / T\)\\
        \hline
        $8$ & -- & $12$ & \(2^{28}\) & \(25.000\)\%  \\ 
        \hline
        $10$ & -- & $15$ & \(2^{35}\) & \(0.1953\)\%  \\ 
        \hline
        $8$ & -- & $16$ & \(2^{32}\) & \(1.5620\)\%  \\ 
        \hline
        $10$ & -- & $20$ & \(2^{40}\) & \(0.0061\)\% \\
        \hline
        $8$ & $16$ & $12$ & \(2^{44}\) & \(0.0004\)\% \\
        \hline
         $10$ & $20$ & $15$ & \(2^{55}\) & \(1.863\cdot 10^{-7}\)\% \\        
        \hline
        $8$ & $16$ & $16$ & \(2^{48}\) & \(2.384\cdot10^{-5}\)\% \\
        \hline
        $10$ & $20$ & $20$ & \(2^{60}\) & \(5.821\cdot 10^{-9}\)\% \\        
        \hline
    \end{tabular}
    \end{center}
\end{table}
In general, we observe improved RDFC performance with both (i) lower mutual information \(I(X;Y)\), which is $\approx 0.5$ and $\approx 0.19$ for $p=0.11$ and $0.25$, respectively, and (ii) higher RDFC rate \(\jrate\), which we expect by (\ref{th:rate-region}). However, \emph{only} increasing \(\jrate\) leads to slightly worse performance. This is likely due to the static sample size \(\numsamp\). We do not adapt \(\numsamp\) to the different alphabet sizes between experiments due to practical memory and time reasons, which can lead to the training set being less representative in some experiments than others. We list the ratios of \(\numsamp\) to all possible combinations \(T = |\xalph||\yalph||\kalph||\lalph|\) in our experiments in Table~\ref{tab:relative-samples}.

\subsection{AE Design Insights}
\label{sec:iv-ae-design}
We observe faster convergence during training when layer 7 in Table~\ref{tab:ae-layout} has the sigmoid activation function and a lower quantization error between the encoder output \(\tilde{\jrealiz}\) and the VQ output \(\jrealiz\) compared to when ReLU is used. We conjecture that the activation function normalizes \(\tilde{\jrealiz}\) to an interval close to the latent alphabet \(\jalph\), which seems to aid in training so that the encoder adapts better to the VQ-layer and the constraints it imposes. Moreover, a deep encoder and decoder converge faster than shallower architectures for a fixed network width. Our simulations do not indicate whether this is solely due to the larger number of network parameters or whether the network depth alone enables more complex behavior. Finally, in Algorithm~\ref{alg:KLBinning}, we observe increased RDFC performance when we allow \(\lbins(\xrealiz, \yrealiz)\!=\!\varnothing\) for $(\xrealiz, \yrealiz,b)$ such that we have \(\Pr[\yrseq\!=\!\yrealiz|\xrseq\!=\!\xrealiz,\ybins_b]  \approx 0\). 
However, allowing empty sets in \(\kbins(\xrealiz)\) shows instances of both improved and reduced performance depending on \(\targetCond\) and \(\blocklen\).

\section{Conclusion}
\label{sec:v-conclusion}

We proposed AE designs for the RDFC framework, providing significant communication load gains for multiple distributed computation and learning problems. We demonstrated that our AE designs achieve high RDFC performance for distributed synthesis of a BSC, where the amount of common randomness was shown to play a crucial role. Furthermore, we provided fundamental information-theoretic insights into the network architecture and the training-data generation. With this work, we aimed to facilitate the use of RDFC techniques for cutting-edge use cases, including neural image compression, FL with side information, and distributed secure and private function computation. Finally, while one can show that there is a gap between the achieved rate tuples and the asymptotic RDFC limits, such a comparison does not provide meaningful insights due to the short blocklengths considered in this work. Thus, in future work, we will apply, for example, hybrid coding methods introduced in \cite{OurConcatenated} to our AE designs to consider practical blocklengths.

\section*{Acknowledgment}

This work has been supported by the ZENITH Research and Leadership Career Development Fund under Grant ID23.01, ELLIIT funding endowed by the Swedish government, and the Swedish Foundation for Strategic Research (SSF) under the Grant ID24-0087. The computations and data handling were enabled by resources provided by the National Supercomputer Centre (NSC), funded by Linköping University.

\bibliographystyle{IEEEtran}
\bibliography{IEEEabrv,bibliography}

\newpage

\vfill

\end{document}